\documentstyle[psfig]{l-aa}

\begin{document}

\thesaurus{11         
           (11.01.2;  
            11.05.2;  
            11.06.2;  
            11.09.2;  
            11.19.3;  
            11.19.7)} 

\title{A statistical study of the spectra of very luminous IRAS Galaxies}
\subtitle{II.Spectral and environmental analysis}

\author{ H. Wu\inst{1}, Z.L. Zou\inst{1}, X.Y. Xia\inst{2,1} and Z.G. Deng\inst{3,1} }

\offprints{to H. Wu}

\institute{Beijing Astronomical Observatory, Chinese Academy of Sciences, Beijing 100080, China 
\and  Department of Physics, Tianjin Normal University, Tianjin 300074, China
\and  Department of Physics, Chinese Academy of Sciences Graduate School, Beijing 100039, China}

\date{}

\maketitle

\begin{abstract}\footnotemark{}
    \footnotetext{Table 1 is only available in electronic 
form at the CDS via anonymous ftp to cdsarc.u-strasbg.fr(130.
79.128.5) or via http: 
//cdsweb.u-strasbg.fr/Abstract.html.}


Spectroscopic observations of a sample of 73 very luminous IRAS galaxies
($\rm log(L_{IR}/L_{\odot})\ge11.5$ for $\rm H_{0}=50km s^{-1} Mpc^{-1}$,$\rm 
q_{0}=0.5$) from the 2Jy redshift survey catalogue were carried out using 
the 2.16m telescope at the
Beijing Astronomical Observatory.  The observational data, including
the optical images (extracted
from Digital Sky Survey) and spectra for these galaxies, are presented
in paper I (Wu et al., 1998).
In this paper, we give the spectral and morphological classifications 
for these very luminous IRAS galaxies (VLIRGs). We show that about
60\% of VLIRGs exhibit AGN-like spectra (Seyfert 1s, Seyfert 2s, LINER-like galaxies).
This fraction goes
up to 82\% for the ultraluminous IRAS galaxies (ULIRGs) subsample 
($\rm Log(L_{IR}/L_{\odot}) \geq 12.0$). 56\% of the VLIRGs show strong interaction
or merging signatures; this fraction rises to 91\% for the ULIRGs.
These statistical results strongly suggest that
interaction triggers nuclear activities and enhances the infrared luminosity. We find that
LINER and a mixture type which have optical properties of both HII galaxies and LINERs
could be at the transition stage from infrared luminous HII galaxies to AGNs;
their main energy production is from starbursts as well as AGNs. 

Both infrared luminosities 
and $\rm H\alpha$ equivalent widths increase dramatically as nuclear 
separations between VLIRGs and their nearest neighbors decrease.
 There is little doubt that strong starbursts happen in the nuclei of VLIRGs.
Assuming class 0 as advanced merger, we construct a simple merger sequence, 
from morphological classes 1 to 4 (with near or far companions), 
to class 5 and 6 (interacting pairs and mergers) and then to class 0 (isolated galaxies). 
Along this sequence, VLIRGs evolve from HII galaxies to AGNs. 


\end{abstract}

\section{ Introduction}

Very luminous infrared galaxies (VLIRGs) , especially ultraluminous infrared
galaxies (ULIRGs) have bolometric luminosities comparable with 
quasars and dominate the top end of the galaxy luminosity function (Soifer, et al., 1987).
Many theoretical and observational studies (Sanders et al., 1988; 
Norris et al., 1988;
Struck-Marcel and Tinsley, 1978; Rieke et al., 1980; Norman and Scoville, 1988;
Rieke, 1988; Condon et al., 1991; Leitherer and Heckman, 1995) show that the high
infrared luminosities are associated with the phenomena of 
active galactic nuclei (AGNs) and starbursts,
rather than the radiation from an old stellar population
(Thronson, et al., 1990) and cloud-cloud collision (Harwit, et al., 1987). 

The main energy output mechanism for luminous infrared galaxies is still
being debated. Both  AGNs and starbursts have
been proposed as the main energy provider. Sanders et al. (1988) found that
a high proportion of AGNs seems to be present in ULIRGs, 
but Rigopoulou, Lawrence and Rowan-Robinson (1996) found that
the starburst model fits well the
radio, sub-millimeter/IR to soft X-ray observations
for Sanders' 10 ULIRGs.
Recent 3D infrared spectroscopic 
imaging observation (Kroker et al., 1996) suggests that both Seyfert 1 nucleus 
and circumnuclear star formation contribute significantly to the total luminosity. 
It is possible that AGNs and starbursts co-exist in the central region.
Recently, Veilleux, Sanders and Kim (1997) reported the  results of a
near-infrared 
search for hidden broad-line regions in a sample of 25 ULIRGs with no
broad-line regions in the optical band.
They detected broad recombination lines in five objects and all of them are Seyfert 2 galaxies.
This observation provides
important clue that there could exist evolution connection from
nuclear starbursts to AGNs in the central region of luminous infrared galaxies.

In order to learn
more about the central region of luminous infrared galaxies, and to
understand the possible evolutionary process from starbursts to AGNs and 
how they are associated 
with galaxy' interaction/merging,
we carried out spectroscopic observations
for a large sample of 73 VLIRGs
compiled from the 2Jy redshift survey catalogue (Strauss et
al. 1990, 1992). The spectra and the optical images were presented in a
companion paper (Wu et al. 1998, hereafter paper I).
In this paper,
we report the spectral classification and analysis for the sample
(section 2). The
morphological classification and their environmental
properties are discussed in section 3.
In section 4, we point out
 a possible spectral evolutionary process of luminous infrared
galaxies using the relationship between the spectral and morphological 
classifications.
 A possible nuclear and evolutionary model for these galaxies
is then constructed.
Throughout this paper, we use 
$\rm H_{0}=50km s^{-1} Mpc^{-1}$ and $\rm 
q_{0}=0.5$.

\section{ Spectral analysis}

\subsection{ Choice of line ratios}

Active
galaxies are normally classified spectroscopically
by several emission line ratios
(Veilleux and Osterbrock 1987, hereafter
VO87). Following VO87,
we select lines with higher signal to noise ratios. The ratios are
formed by only one element or with HI 
Balmer lines and the wavelength separation
are chosen to be small,
in order to minimize the effects caused by
low accuracy, reddening, flux calibration and different metallic abundances.
The ratios we used are $\rm [OIII] \lambda 5007 / H\beta$, $\rm [NII] \lambda 6584 / H\alpha$,  
$\rm [SII](\lambda 6716+\lambda 6731)/H\alpha$ and $\rm [OI]\lambda6300/H\alpha$.

\begin{figure}
\centerline{\psfig{figure=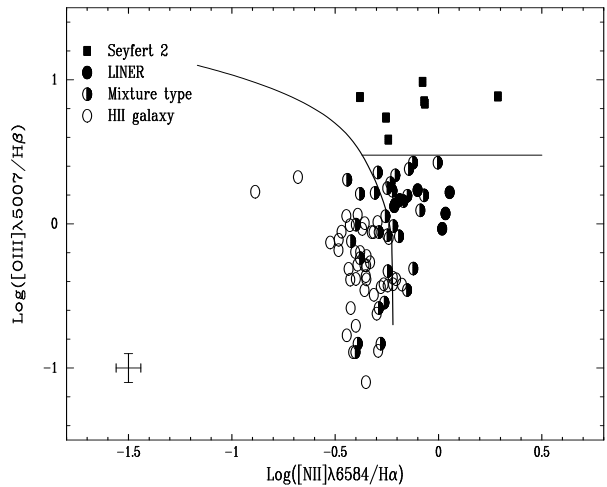,height=6.50cm,width=8.0cm,angle=270}}
\caption[1a]{De-reddened flux ratio $\rm [NII]\lambda6584/H\alpha$ as a function
of $\rm [OIII]\lambda5007/H\beta$ for VLIRGs. The curve and horizontal
line are from VO87.
In the low left of the figure, the mean error bar is indicated.
}
\label{fig1}
\end{figure}

\begin{figure}
\centerline{\psfig{figure=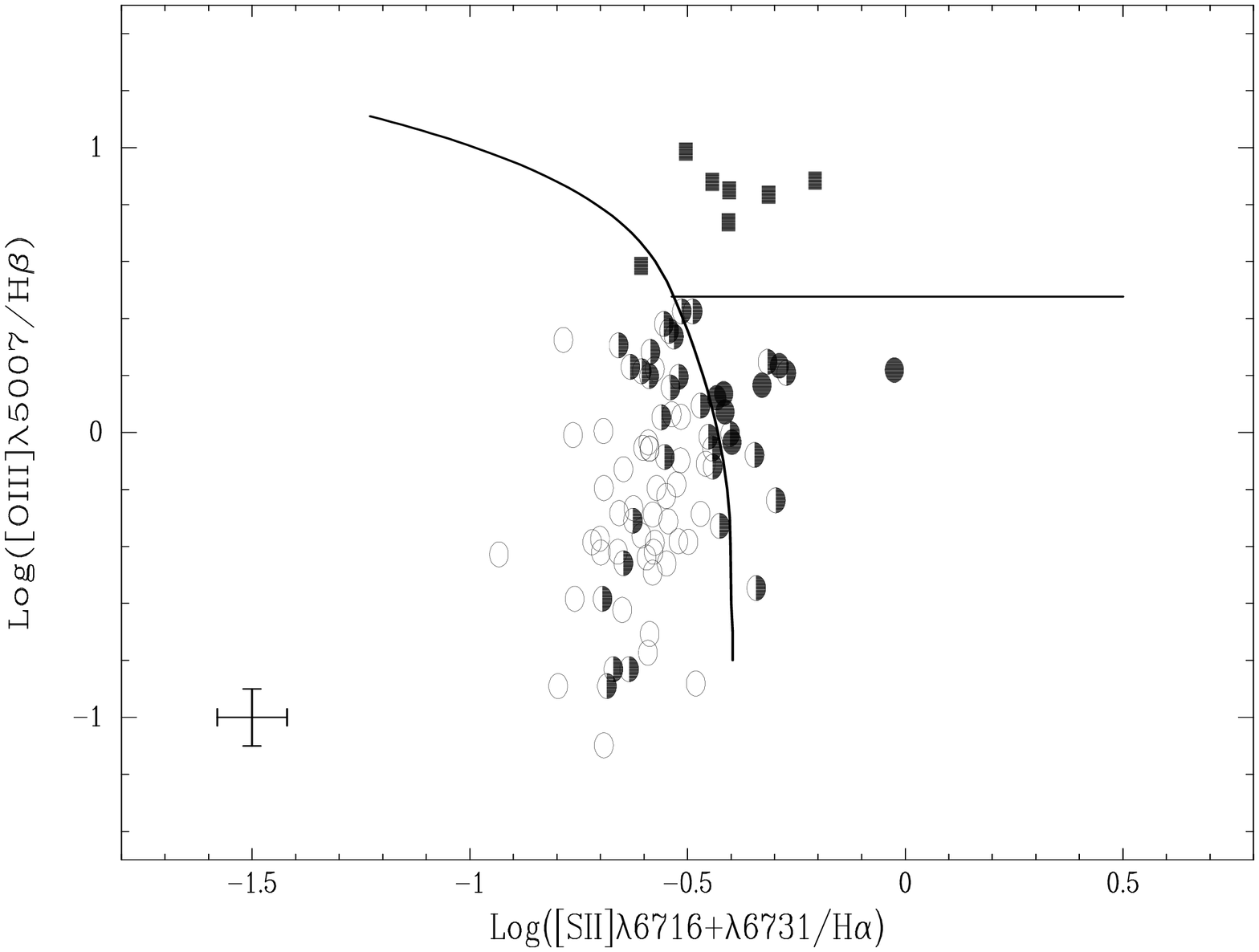,height=6.50cm,width=8.0cm,angle=270}}
\caption{De-reddened flux ratio $\rm [SII]\lambda6716+\lambda6731/H\alpha$
as a function of $\rm [OIII]\lambda5007/H\beta$ for VLIRGs. The meanings 
of symbols and curves are the same as in Fig.1.
}
\label{fig2}
\end{figure}

\begin{figure}
\centerline{\psfig{figure=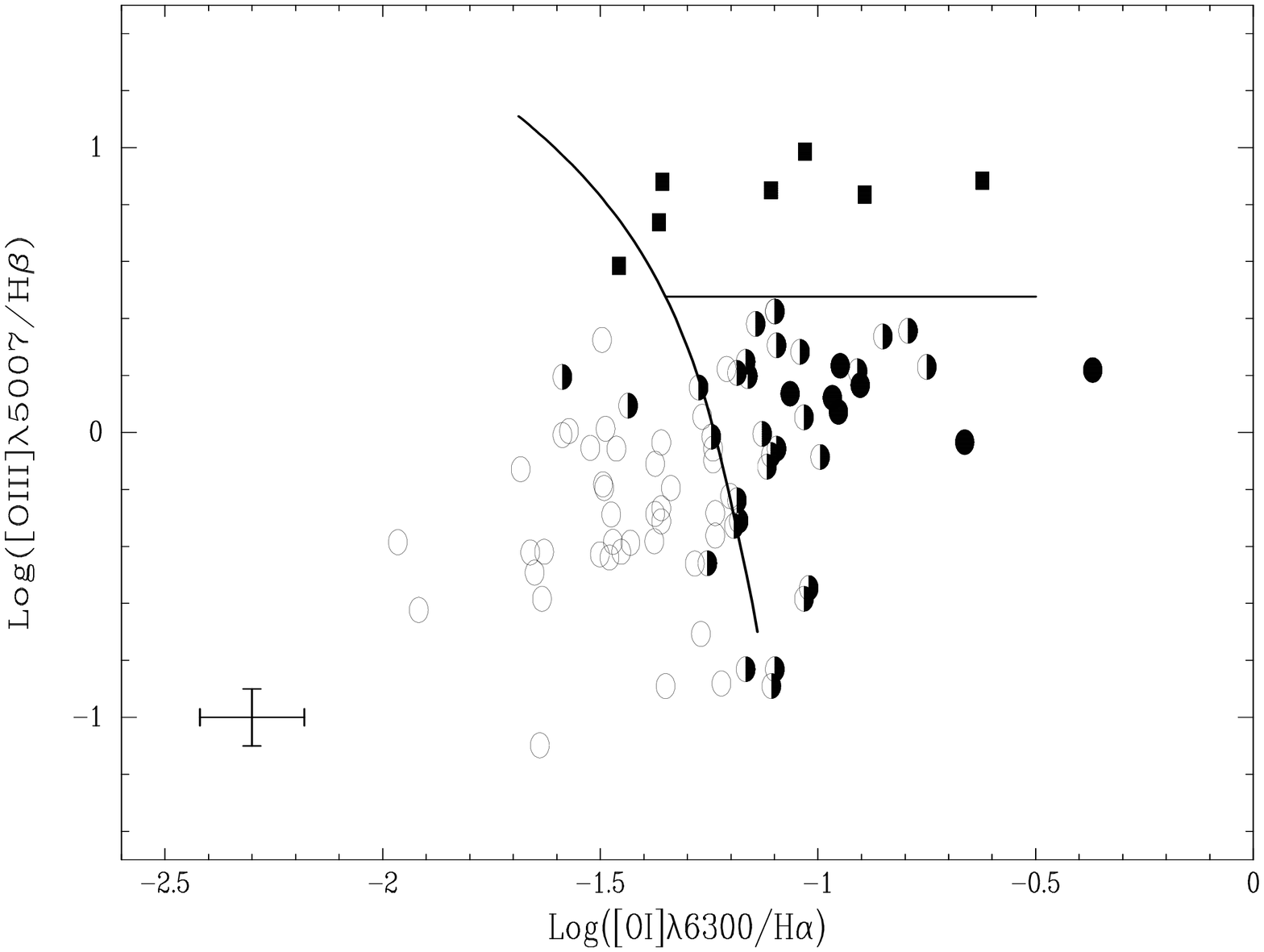,height=6.50cm,width=8.0cm,angle=270}}
\caption{De-reddened flux ratio $\rm [OI]\lambda6300/H\alpha$ as a function
of $\rm [OIII]\lambda5007/H\beta$ for VLIRGs. The meanings of symbols and 
curves are the same as in Fig.1.
}
\label{fig3}
\end{figure}

\subsection{Line ratios}

Before performing spectral classifications, we first
correct reddening for emission lines.
In principle, reddening includes contributions from the Galaxy, the
intergalactic medium and the host galaxy. We ignore the second term
since it is expected to be small and there is little knowledge.
The Galactic reddening
is corrected according to the Burstein and Heiles' (1984) machine-readable reddening table.
Since almost all the galaxies we observed have emission lines, and most of them show both
$\rm H\alpha$ and $\rm H\beta$ emissions, we can obtain the 
host galaxy reddening from the ratios of $\rm H\alpha/H\beta$.
The Whitford reddening curve as parameterized by Miller and Mathews (1972) is used. We adopt an
intrinsic $\rm H\alpha/H\beta$ ratio of 2.85 for HII galaxies and 3.10 for LINERs and Seyferts 
(Veilleux et al., 1995, hereafter V95).
The color excesses E(B-V) can be derived by the 
relation $\rm E(B-V)=0.77E_{\beta-\alpha}$ (Gebel, 1968 and VO87), here
$\rm E_{\beta-\alpha}$ is defined by (Miller and Mathews, 1972):
$$
\rm E_{\beta-\alpha} = 2.5log[(I(H\beta)/I(H\alpha))_{theory}/(I(H\beta)/I(H\alpha))_{obs}].
$$

The Galactic color excess, observed and theoretical $\rm I(H\alpha)/I(H\beta)$, 
color excess of host galaxies, observed and de-reddened emission line ratios are
listed from column 2 to column 8 in table 1. We also give the continuum ratios
at $\rm 6563\AA$ and $\rm 4861\AA$ in column 9. The  uncertainty is about
15\% as discussed in paper I.  Colons(:) and semicolons(;) indicate values with 
uncertainties about 30\% and 50\% respectively. 

It is clear from table 1 that for most of our sample galaxies, the reddening corrections 
are very small, so the errors mainly come from the measurement and aperture 
effects due to different slit-widths and extraction windows.
 Since most of the sources have strong line emission 
of $\rm H\alpha$,$\rm H\beta$, $\rm [NII]\lambda6584$, $\rm [SII]\lambda6716+\lambda6731$ 
and $\rm [OIII]\lambda5007$, the errors from measurements are less than 10\%. 
From the observation and data reduction, we found that aperture effect
introduce errors no larger than the measurement errors. Therefore the final errors 
for these lines are about 10\%-15\%.  For the $\rm [OI]\lambda6300$ line
the lower S/N and possible blending make the measurement harder,
as a result the line ratio $\rm [OI]\lambda6300/H\alpha$ is more uncertain
(cf. paper I and section 4.1).

\subsection{Spectral classification}

The early classification criterion was given by Baldwin, Phillips and Terlevich (1981),
which was based on the ratio of $\rm [OII]\lambda3727 / [OIII]\lambda5007$ and mean excitation
to distinguish 
HII galaxies, Seyfert 2s and LINERs. However,
most of our spectra do not cover the $\rm [OII]\lambda3727$ line and further
a reliable determination of this line ratio is often difficult even when
this line is covered due to its low S/N,
therefore, following Osterbrock (1989), we use the line ratios 
of $\rm [OIII]\lambda5007/H\beta$, $\rm [NII]\lambda6584/H\alpha$,
$\rm [OI]\lambda6300/H\alpha$ and $\rm
[SII](\lambda6716+\lambda6731)/H\alpha$ for classification purposes instead.
The diagnostic diagrams of classifications are shown in Figs. 1-3.
The curve and horizontal line at figures divide each panel into three areas.
HII galaxies locate at the left side of the curve.
The line $\rm [OIII]\lambda5007/H\beta=3$,
is used to separate AGN with high excitations --
Seyfert 2's from LINERs. 
We also plot the mean error bar in the low-left.

It is obvious from Figs.1-3 that there is no clear boundary to separate
the different types of galaxies, especially HII galaxies and LINERs. 
There are quite a few sample galaxies located near the solid curves at figures.
Though we have considered the possible errors, there are still some objects 
which can not be satisfy all the definitions of either
HII galaxies or LINERs; for these galaxies they locate at HII galaxies
region in one or two diagnostic diagrams, but locate at LINERs region in other
diagnostic diagrams.
We therefore
classify these galaxies as a mixture type. The mixture type
galaxies show some similar properties as the LINERs and could be a
transitional phase from HII galaxies to AGN (see section 4).

  It is clear from Figs. 1-3 that there are no galaxies with relatively high ionization
level
($\rm [OIII]\lambda5007/H\beta > 3$) at the 
HII galaxy region. This distribution 
of VLIRGs at the diagnostic diagrams is quite different from the optical
sample of VO87, but similar with the results from the
the IRAS Bright Galaxy Survey (BGS, V95)
and the infrared color-selected samples (Armus, Heckman and Miley,
1989). As pointed out by Allen et al. (1991) that the high ionization
HII galaxies 
include extreme starbursts and WR galaxies, the lack of high ionization
galaxies in 
VLIRGs (We found only few WR galaxy in our sample)
indicates that VLIRGs do not 
belong to the young starburst population. 

The classifications are listed in column 10 of table 1. ``H" stands for
HII galaxy, 
``L" for LINER, ``S2" for Seyfert 2 and ``S1" for
 Seyfert 1. We use ``LH" to represent 
the mixture type with properties of both HII galaxies and LINERs. 
``S?" is for unclassified AGNs
which can only be distinguished from HII galaxy, due to the lack of 
data on $\rm [OIII]\lambda5007/H\beta$.
"O" represents galaxies which can not be classified because
they show no or few observed emission lines.
We also list the Verons' classification (1993) for some sources in column 11. 
For most objects the two classifications are consistent. However, there are two
objects IR23254+0830 and IR13536+1836 which are classified as Seyfert
1's
 in Verons'
Catalogue but show Seyfert 2 properties in our optical spectra 
and show Seyfert 1
properties only in polarized spectra (Miller and Goodrich, 1990). 
A Seyfert 1 galaxy -- IR23532+2513B 
(Zou et al., 1995) and a Seyfert 2 galaxy --IR23594+3622 are
discovered by us.

\begin{figure*}[t]
\centerline{\psfig{figure=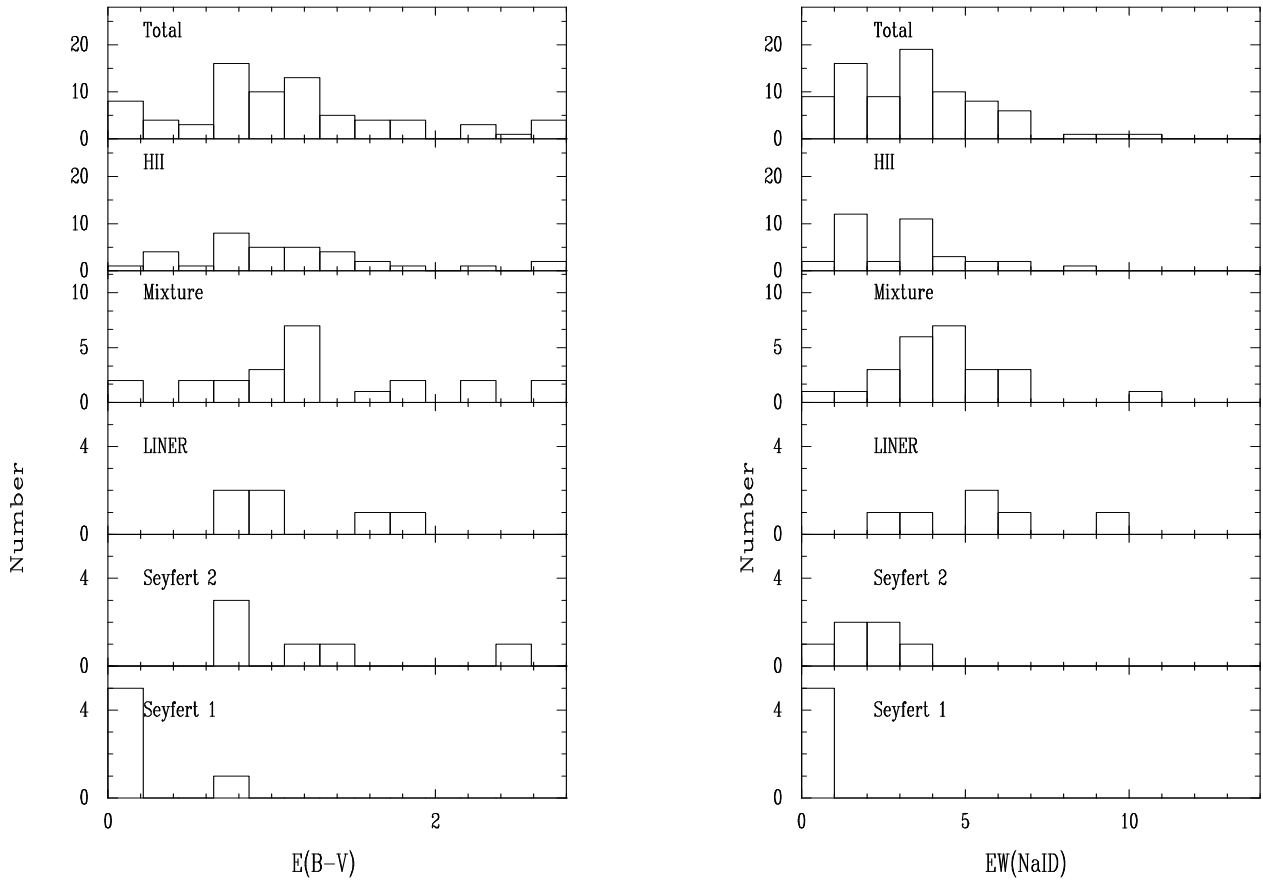,height=12.00cm,width=17.0cm,angle=270}}
\caption{The distributions of color excess and equivalent widths
for the NaID line for different spectral types.
}
\label{fig4}
\end{figure*}

\subsection{Statistic results}

Because more than 40\% of our sample galaxies are in pairs, possible groups of
galaxies or the systems with double nuclei or multi-nuclei of galaxy,
the first
thing is to determine which component emits the high infrared luminosity.
For most target galaxies, the identification is easy
using the IRAS position
and error ellipse.
But for a few sources, we  have to use the infrared colors 
or follow the method
of V95 by orders of activity (Seyefrt 1, Seyfert 2, LINER and HII galaxy)  
to identify the optical counterparts.

The statistical results of classification are listed in table 2a-c.
We present the results of both spectral and morphological classes in these tables.
Morphological class would be discussed in section 3.1.
In order to keep the results of statistics complete, throughout the paper
we use the sample of 73 VLIRGs for statistics.
   From table 2, we can see that there are 40\% (29/73) HII galaxies in our VLIRGs sample,
which is lower than 59\% in BGS (V95). It results from our selection
of IRAS galaxies with higher
infrared luminosity. AGNs (Seyfert 1s, Seyfert 2s, LINERs
and mixture types) appear in 44 nuclei (60\%) of galaxies.
Here we include the mixture types,
 because they show some similar properties as LINERs
as will be shown in section 4.2.
There are 6 Seyfert 1's (8\%), 6 Seyfert 2's (8\%), 6 LINERs (8\%), 
23 mixture types (32\%) and 3 unclassified AGNs (4\%).
 It is obvious that LINER-like 
objects (LINERs and mixture types)
dominate the infrared luminous AGN (66\%, 29/44).
For a ultraluminous subsample with 11 galaxies, 
there are 2 HII galaxies (18\%), 3 Seyfert 1's (27\%), 1 Seyfert 2's (9\%),
2 LINERs (18\%), 2 mixture types (18\%) and 1 unclassified AGN (9\%).
For the subsample with
$\rm 11.5\leq log(L_{IR}/L_{\odot})\leq 12.0$, there are 3 Seyfert 1s (5\%), 
5 Seyfert 2s (8\%), 4 LINERs (6\%), 21 mixtures (34\%), 2 unclassified AGNs (3\%) and 
27 HII galaxies (44\%).

It is clear from our statistical results that the proportion of AGNs (82\%) 
in the ultraluminous subsample is much higher than that in relatively lower 
luminous subsample (56\%). 
In addition, the LINER-like galaxies in AGNs decrease 
rapidly from 71\% (25/35) to 44\% (4/9), in contrast
Seyfert 1s in AGNs increase 
from 9\% (3/35) to 33\% (3/9). All these results support the point of view 
of V95 that the proportions of AGN-like and Seyfert-like galaxies 
increase as the infrared luminosity increases. The proportions of
Seyferts are 
13\% (8/62) and 36\% (4/11) for our lower and higher infrared luminosities' 
subsamples, respectively. We conclude that our two subsamples have
different statistical properties;
higher infrared luminosity relates strongly with AGN phenomena. 

\subsection{Dust in VLIRGs}

\begin{figure}
\centerline{\psfig{figure=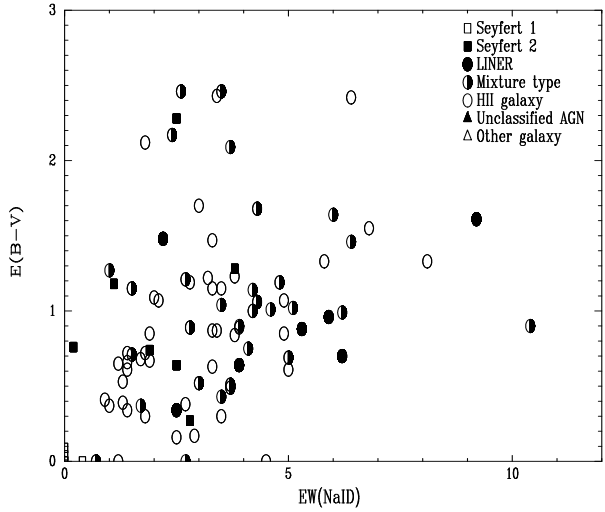,height=6.50cm,width=8.0cm,angle=270}}
\caption{Color excess as a function of the equivalent width of the
NaID line for each spectral type. The meaning of each symbol is shown in
the up-right of the plot. In the following figures, all the symbols have
the same meanings.
}
\label{fig5}
\end{figure}

\begin{figure}
\centerline{\psfig{figure=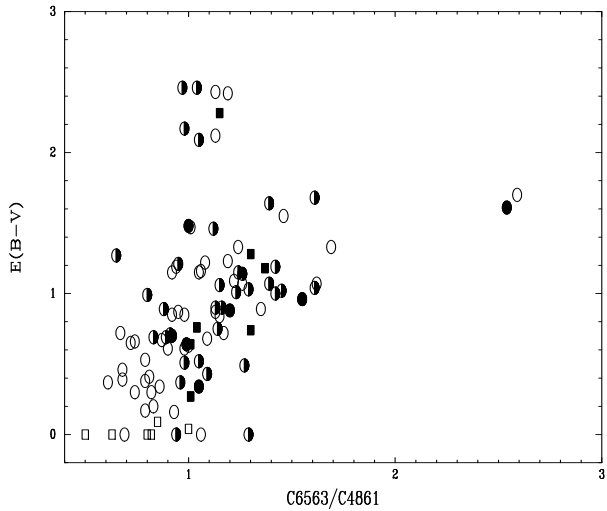,height=6.50cm,width=8.0cm,angle=270}}
\caption{Color excess as a function of the observed ratio of the
continuum at 6563$\rm \AA$ and 4861$\rm \AA$ for each spectral type. 
}
\label{fig6}
\end{figure}

\begin{figure}
\centerline{\psfig{figure=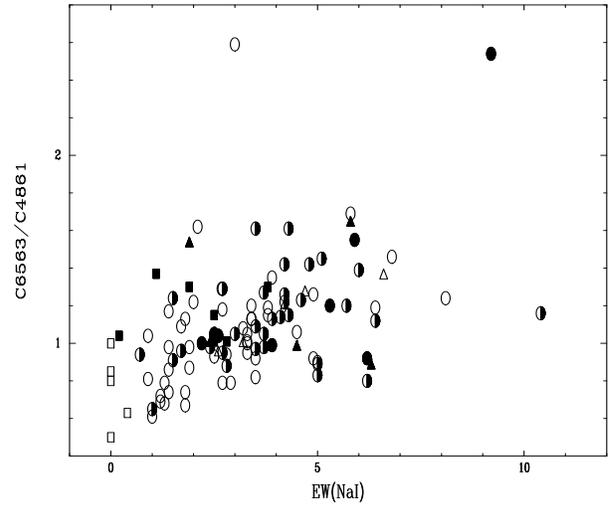,height=6.50cm,width=8.0cm,angle=270}}
\caption{Observed continuum color ratio C6563/C4861 as a function of the 
equivalent width of NaID for each spectral type.
}
\label{fig7}
\end{figure}
\begin{figure}
\centerline{\psfig{figure=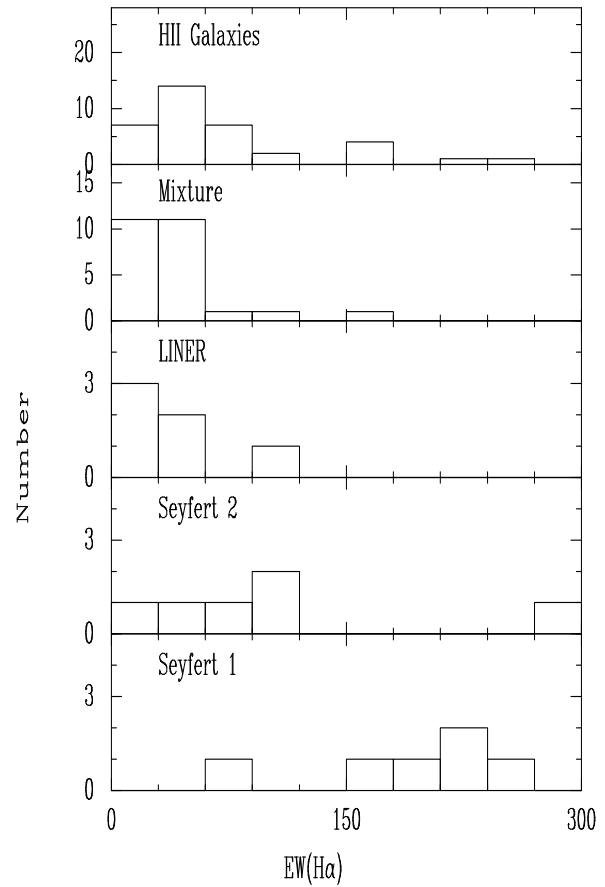,height=12.00cm,width=8.0cm,angle=270}}
\caption{Distributions of the equivalent widths for $\rm H\alpha$
of each spectral type.
}
\label{fig8}
\end{figure}

We have three dust indicators: E(B-V), EW(NaID) and C6563/C4861(Ratio of continuum fluxes
at 6563$\rm \AA$ and 4861$\rm \AA$). 
E(B-V) represents the reddening of line emission region,
C6563/C4861 probes the reddening of continuum emission region and EW(NaID)
the interstellar reddening. 
E(B-V) is obtained from the ratio of $\rm H\alpha/H\beta$. 
The systematic errors in E(B-V) is unavoidable from the larger 
wavelength separation of $\rm H\alpha$ and $\rm H\beta$. Furthermore,
the aperture sizes in observation and extraction could introduce uncertainty 
due to the  complex geometric  structure of dusty regions
in the nuclei of VLIRGs. 
The underlying stellar absorption of $\rm H\alpha$ and $\rm H\beta$ could 
also influences the accurate measurement of $\rm H\alpha$ and $\rm H\beta$ emission.
Combined all these possible factors, the uncertainty of $\rm H\alpha/H\beta$
could be as large as 20\%-25\%. Similar to $\rm H\alpha/H\beta$, the measurement 
of continuum C6563/C4861 may also have a larger uncertainty, $\sim$ 20\%, 
because of the large wavelength separation.
For EW(NaID), the error is about 10\%-15\%.

Fig. 4 is a histogram of E(B-V) for different types of VLIRGs. The median values 
of E(B-V) for Seyfert 1s, Seyfert 2s, LINERs, Mixtures and HII galaxies 
are 0.02, 0.97, 0.92, 1.02 and 0.86 respectively.
K-S tests show that the probability that Seyfert 1 are drawn from the same distribution
as the other four types of galaxies is less than 0.03. 
Fig. 4 also shows the distributions of EW(NaID). It shows that
the Seyfert galaxies have lower median EW(NaID) (0.0$\rm\AA$ for Seyfert 1's
and 2.2$\rm\AA$ for Seyfert 2's) than that of others (3.2$\rm\AA$ 
for HII galaxies, 4.2$\rm\AA$ for mixture types and 5.6$\rm\AA$ for LINERs).
LINERs and mixture types have larger NaID absorption than HII galaxies.
Both E(B-V) and EW(NaID) show that Seyfert 1s have less dust
and LINER-like galaxies show more dust
than that of HII galaxies.  

Fig. 5 is a plot of EW(NaID) versus E(B-V) for different types of VLIRGs. 
Seyfert 1's have the lowest E(B-V) and EW(NaID) values, indicating that
they are not much affected by dust.  For HII galaxies,
there is a significant correlation between the E(B-V) and EW(NaID);
the probability that the correlation is fortuitous is 0.015.
supports the interstellar origin of NaID absorption.  
Similar correlations exist among other dust indicators for HII galaxies (Figs. 6 and 7):
the probability of no-correlation between C6563/C4861 
and E(B-V) for HII galaxies is 0.0016, and that between C6563/C4861 and EW(NaID) is 0.02.
These strong correlations indicate that all three reddenings have a similar origin
and HII galaxies have a simple nuclear structure. 

However the statistical tests do not show significant correlations among 
the three reddening indicators for the LINER-like galaxies and Seyfert 2's. 
This indicates that the line and continuum emission may come from different
regions in the nuclei, and the nuclear or circumnuclear regions are 
much complex compared with those of HII galaxies.

\subsection{$\rm H\alpha$ emission}

As an indicator of star formation, $\rm H\alpha$ emission is clearly
quite important.
Fig. 8 shows the distribution of equivalent width of $\rm H\alpha$
for different types of galaxies.
The median EW($\rm H\alpha$) are 197$\rm\AA$, 89$\rm\AA$, 35$\rm\AA$, 34$\rm\AA$ 
and 55$\rm\AA$ for Seyfert 1s, Seyfert 2s, LINERs, mixture types and HII galaxies 
respectively. It is clear that the Seyfert galaxies have the
strongest $\rm H{\alpha}$ emission, and LINER-like galaxies  the weakest. 
We found that the LINER and mixture type of galaxies have similar median value 
of EW($\rm H\alpha$). K-S tests show that, among the five types,
only LINERs and mixture types
could come from the same population (the K-S probability is 0.996).
It indicates that
these two types have similar $\rm H\alpha$ emission properties. 
therefore, it is reasonable to combine them as a LINER-like type
as we have used in section 2.4 (see also section 4.2).

\section{Galaxy interaction}
\subsection{Classification of galaxy interactions}

A number of CCD images of ULIRGs are now available,
 such as deep optical images (Leech et al.
1994), near-infrared images (Murphy et al., 1996) 
and HST images (Surace et al., 1998). Unfortunately, 
the overlap number of galaxies with CCD images
with our sample is small. 
Our morphological classification and the study of environments of VLIRGs 
are based on the Digital Sky Survey (DSS) images (see Paper I). For the
 overlap galaxies with CCD images,
our classification
 do not conflict with those based on the near-infrared or 
HST images. 

We use
the classification scheme for VLIRGs of Lawrence et al. (1989), who
define isolated galaxies as class 0, galaxies with far and near companions 
as class 1 to 4, interacting pairs
as class 5  and mergers as class 6.
Morphological classifications and environment parameters for our sample
galaxies are listed in table 3.  In practice,
we find it difficult to separate class 5
from class 6, because there is no exact dividing line
between interacting pairs and mergers,
it is therefore better to join classes 5 and 6 into a single class.  
Also one
should be careful to define class 0. Our statistics show that this type
has similar maximum infrared luminosity to that of class 5 and 6 
and higher than that of class 1-4. 
Moreover, the percentage of AGN in class 0 is even higher
than that in class 5 and 6 (see section 4.3).
Therefore, it is likely
that the class 0 is in the advanced or post merging stages.

As described in section 2, table 2a-2c
list the statistics of morphological types
for the whole sample and two subsamples.
It is clear from table 2a that 56\% (41/73) of VLIRGs are 
interacting or merging systems (class 5 or class 6). 18\% (13/73)
are isolated (class 0). The fraction of class 1-4 objects are about 26\%. 

For the subsample with $\rm Log(L_{IR}/L_{\odot}) \geq 12.0$, 
nearly all the objects, namely 91\%  of ULIRGs,
are strong interaction or merging systems 
(there are 10 ULIRGs belong to class 6 and
the rest belong to class 0). 
These  results are consistent with those of Sanders et al. (1988),
Melnick and Mirabel(1990) and Clements et al. (1996a,b); the fractions are,
however, higher
than those found by Zou et al. (1991, 1993) and Leech et al. (1994).
The different results for ULIRGs can be understood as 
follows:
most of ULIRGs in Sanders', Melnick's and our sample are nearby galaxies 
with redshifts $\rm z < 0.13 $,
for which the disturbed features could be detected 
more easily than the ULIRGs in Zou et al.'s and Leech et al.'s samples,
which are most far-away.

In order to avoid the large optical reddening, Murphy et al. (1996) obtained
near-infrared images of 56 ULIRGs, 95\% of these
show evidence for current or past 
interaction. Recently, Surace et al.(1998) present HST images of 9 ULIRGs,
in which 8 sources are advanced mergers. These near-infrared and
high-resolution HST images show that
there may exist very
few truly isolated ULIRGs with no
disturbed features. 
It should be very interesting  
to study the small class 
of ``isolated'' galaxies with high resolution and/or infrared
imagings.

It is also obvious from table 3 and DSS images that at 
least 14\% (10/73) of VLIRGs are in groups 
of galaxies or in multiple-merging systems: 
7 VLIRGs are located in 
groups of galaxies which
include at least 3 galaxies with consistent redshifts, 
and 3 VLIRGs
with multiple-nuclei.
For example, IRAS 23532+2513 (Zou, et al., 1995) is in a 
compact group with one Seyfert 1 galaxy and one starburst galaxy.
Mrk 273 is a multiple-merging system
with $\sim$ 10 faint objects (within a projected distance
of $\sim$ 100kpc), one of which is a companion dwarf galaxy
showing an unusually high soft X-ray luminosity (Xia et al., 1998). 
ULIRGs and VLIRGs
seem to be evolutionally connected with compact groups of galaxies;
studying this class
of galaxies may provide important clues for galaxy evolution.

\subsection{Projected separation}

The separation between an infrared galaxy and its nearest neighbor
in double or multiple-nuclei is one important parameter
that describes the intensity of galaxy interaction. It is also
a good indicator of evolution in interacting galaxies. Since
we can not obtain the physical separation
between two objects
we have to use projected separation instead. Bushouse, Lamb and Werner
(1988), Telesco et al. (1988) and Wei (1990) have studied this
parameter on optically
selected interacting pairs. Wei (1994) also extended this to
far-infrared galaxies. But since there was no
redshift data for each source, these results are prone to errors induced
by chance alignment. In contrast, the redshift information obtained from
our observed spectra completely eliminate this error and allow us
to study the relation between projected separations and activities of
VLIRGs.

\begin{figure}
\centerline{\psfig{figure=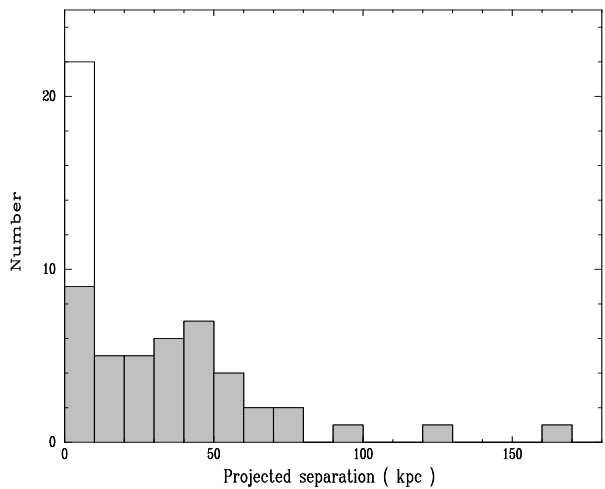,height=6.50cm,width=8.0cm,angle=270}}
\caption{Distribution of projected separations between VLIRG and their nearest
companions. The hollow
histogram indicates the isolated VLIRGs, which could be advanced
mergers.
}
\label{fig9}
\end{figure}

The angular separations measured from the DSS images are
transformed to projected separations using the redshift of infrared galaxies.
For groups or multiple-nuclei, the
minimum separation is used. Fig. 9 shows the distribution of projected
separation. The hollow box includes the isolated galaxies, which are assumed as
advanced merger with only one remaining nucleus. It can be seen that
most VLIRGs have companion within 50kpc.
This is consistent with previous works by several authors that 
more active far-infrared galaxies have 
closer companions.
If we include the isolated galaxies, we find that for large number of VLIRGs 
the project separations are smaller than 10kpc.

\begin{figure}
\centerline{\psfig{figure=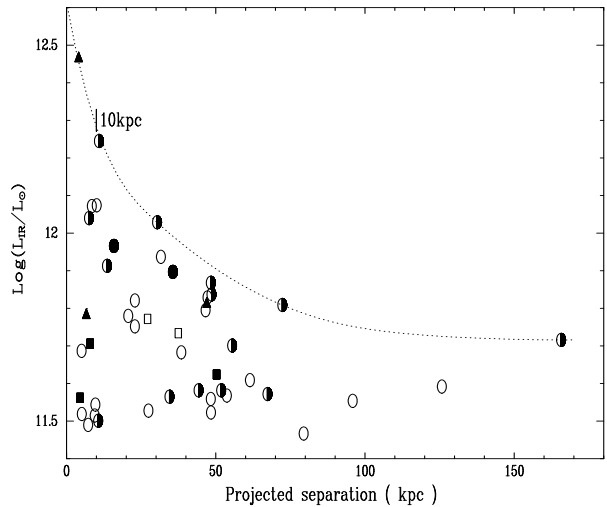,height=6.50cm,width=8.0cm,angle=270}}
\caption{Infrared luminosities as a function of projected separation. 
Dotted line is the upper envelope of data. 10kpc is the character separation
for an IRAS galaxy to be ultraluminous in the infrared.
}
\label{fig10}
\end{figure}

Fig. 10 shows the relation between infrared luminosities and the projected
separations of VLIRGs. Though there is no obvious correlation between them,
the upper limiting boundary line (dotted line)
decreases dramatically as projected separations increase. 
This trend can be easily
understood as a result of projection effects.
It is
reasonable to assume that the fitted envelope line represents
the relationship between infrared luminosity and true separation. 
We define a character separation as the separation when
the infrared luminosity is about half of the peak
value of very close sources (separation $\rm \sim 0$ ). The
separation is found to be 
about 10kpc with a corresponding infrared luminosity
is $\rm 2\times10^{12} L_{\odot}$.
This result
agrees with that of Melnick and Mirabel (1990) and Murphy et al. (1996).

\begin{figure}
\centerline{\psfig{figure=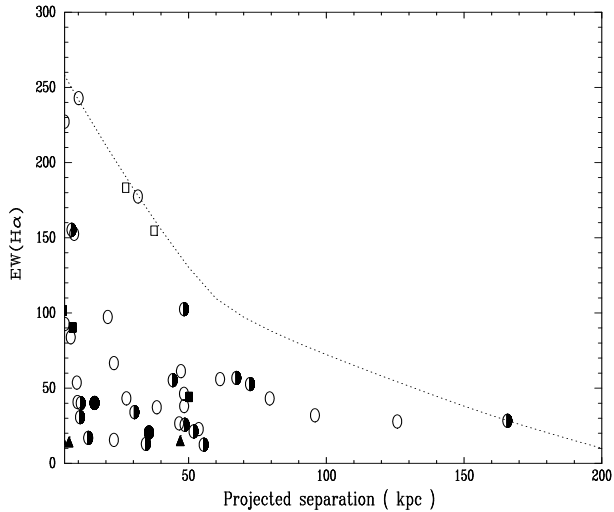,height=6.50cm,width=8.0cm,angle=270}}
\caption{$\rm H\alpha$ equivalent width vs. projected separation.
The dotted line is the upper envelope of data.
}
\label{fig11}
\end{figure}

A similar behavior can be also seen in a
plot of $\rm H\alpha$ equivalent width versus
projected separation (Fig. 11).
The envelope line goes up rapidly as the separation decreases.
Considering that both infrared luminosity and $\rm H\alpha$ 
equivalent width are good
indicators of star formation, Figs. 10 and 11 strongly
suggest that interactions trigger starbursts.

\subsection{Relative Radial velocity}

Besides the projected separation, the relative velocity is another important
parameter that quantifies the interaction between galaxies. 
Since the true relative velocity can not be obtained, we use
the relative radial velocity inferred from  the
redshift difference between an infrared source and
its nearest companion.
The uncertainty in the relative radial velocity
is 30km/s in our data.
Fig. 12 shows the distribution of relative radial velocity.
It is clear that the companions for most VLIRGs have
small relative radial velocities.
Fig. 13 shows that the $\rm H\alpha$ equivalent width
 increases as the relative radial 
velocity decreases. 
This relation is very interesting because it imply that small
relative radial velocity among interaction galaxies should be
the favorate condition for triggering starburst.

\begin{figure}
\centerline{\psfig{figure=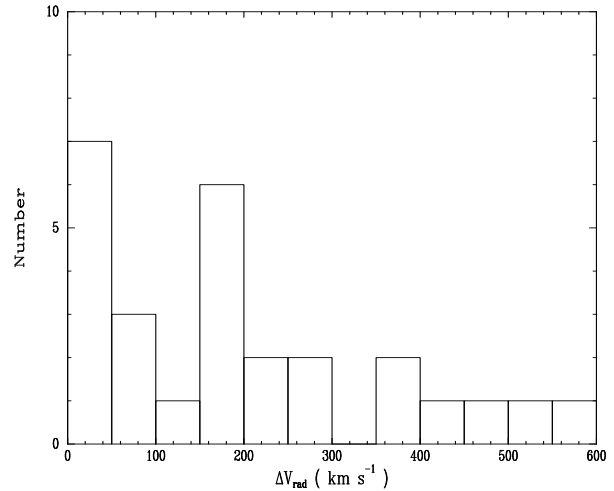,height=6.50cm,width=8.0cm,angle=270}}
\caption{The distribution of relative radial velocities of VLIRGs.
}
\label{fig12}
\end{figure}

\begin{figure}
\centerline{\psfig{figure=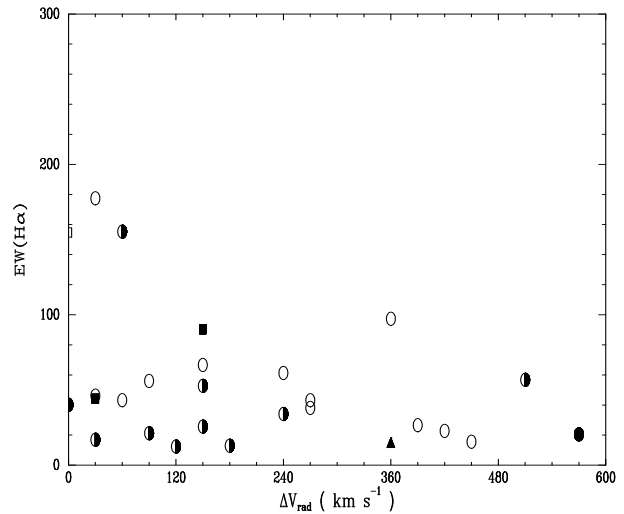,height=6.5cm,width=8.0cm,angle=270}}
\caption{$\rm H\alpha$ equivalent widths vs. relative radial velocities.
}
\label{fig13}
\end{figure}

\section{\bf Discussion}

\subsection{Comparison with V95}

Because our VLIRGs sample are selected from 2Jy redshift survey
catalogue which covers the BGS ($\rm f_{60{\mu}m}\ge5Jy$) sample, there
are a large number of overlap galaxies between our sample and 
the V95 sample. The total overlap number is
37 if we count companions. We compare these overlap galaxies
in the diagnostic emission line ratios, such as 
$\rm [OIII]\lambda5007 / H\beta$,
$\rm [OI]\lambda6300 / H\alpha$, $\rm [NII]\lambda6584 / H\alpha$ and 
$\rm [SII](\lambda6716 + \lambda6731) / H\alpha$.

The ratios of $\rm [NII]\lambda6584/H\alpha$ and 
$\rm [SII]\lambda6716+\lambda6731/H\alpha$ for these two samples
agree very well, with mean difference less than 0.02 and scatter about
0.1;
the difference is within our error bar (see section 2.2).
However,
the mean difference and scatter for the ratios of $\rm [OI]\lambda6300/H\alpha$
are much larger, about 0.06 and 0.21, respectively. The reason is that 
the $\rm [OI]\lambda6300$ emission lines for most sample galaxies are
weak and blended, and are therefore more uncertain.
For $\rm [OIII]\lambda5007/H\beta$, 
the difference and scatter are also large, about 0.05 and 0.23, respectively.
This can not be solely attributed to the low S/N because the 
$\rm [OIII]\lambda5007$ line for most sample galaxies is strong.
As we have discussed before, the reason could be due to
several factors, such as aperture effects, 
underlying stellar absorption of the host galaxy, etc. 

There is a difference in the color excess E(B-V) between
two samples. Our mean value of E(B-V) is 0.99, which is about 0.13 lower than that of
1.12 of V95 in the overlay objects. The Galactic extinction can account for one-fourth(0.03).
The remainder could come from 
different estimates of underlying stellar absorption, 
aperture effect (e.g.,
IR23488+1949A) and possible low S/N of $\rm H\beta$ of spectra.

Since spectral classification is based on emission line ratios, 
aperture effect and other uncertainties can also render
different spectral classifications.
A comparison shows that the our and  V95's classifications agree
for about 70\% (26/37) of the overlap objects. 

\subsection{LINER-like VLIRGs}

The original definition of LINERs by Heckman (1980) was based on the following
conditions on oxygen forbidden line 
ratios, namely,
$\rm [OII]\lambda3727 / [OIII]\lambda5007 > 1$ and 
$\rm [OI]\lambda6300 / [OII]\lambda5007 > 0.33$. However, the spectra
for most of our
galaxies do not cover the wavelength of $\rm [OII]\lambda3727$ and 
the large
wavelength separation between $\rm [OII]\lambda3727$ and
$\rm [OIII]\lambda5007$ could 
bring serious errors due to uncertain
reddening correction. Therefore we use the criteria suggested by
VO87 instead. 
In fact, for the galaxies with $\rm [OIII]\lambda5007$ line, the two definitions
are consistent.

As discussed at section 2.3, some of our sources present properties of both HII galaxy 
and LINER, and could not be classified accurately
as either type. We classify them into a 
mixture type. 
From the spectral analysis, we find that they are similar with LINERs for 
the properties of reddening and $\rm H\alpha$ emission,
so we regard the mixture types 
as LINER-like galaxies. 

Considering that the mixture type VLIRGs show spectral features of
both LINERs and HII galaxies, we can also assume that LINER-like galaxies could be 
the composite systems consisting of both AGN and HII region components. As discussed 
in section 2.3, a few HII VLIRGs present high ionized level, the combination of such low 
ionized HII spectra and AGN spectra is similar to the spectra 
of LINER-like galaxies. Nearby galaxy NGC7679 (V95) is a good example 
for this assumption. As the aperture size increases, the spectrum of NGC7679 changes 
from that of Seyfert 2 to LINER-like type. This supports the idea that NGC7679 has an active nucleus 
and circumnuclear HII region. Recent high resolution HST images (Surace et al. 1998) 
show that there indeed exist many blue ``knots'' of star formation region in the inner 
few kpc of ULIRGs as well as possible active nuclear.

In summary, it is very probably from our spectral analysis that the LINER-like VLIRGs
are mixture type galaxies with both active nucleus and starburst regions. Also
this type of galaxies may at the evolution stage from nuclear starbursts
to AGNs.
As AGN features become
 more dominant, they evolve toward the optical AGNs.

\subsection{Relationship between spectral and interaction classifications}

Spectral classifications provide information about the main energy output
mechanism for galaxies whereas morphological classifications
give the knowledge of evolution phase of interaction galaxies. The correlation
between these two types of classifications may give important clue for understanding
the possible evolution connection between starbursts to AGNs.

In section 3.2 we analyzed
the statistical properties of optical and infrared properties 
as a function of the nuclear separation 
in the double or multiple-nuclear system. As the separation 
decreases, both infrared luminosity 
and  $\rm H\alpha$ equivalent width increase. Since both
infrared luminosity 
and $\rm H\alpha$ equivalent width are indicators of star formation,
it is very clear 
that as the separation decreases, the star formation rate increases.

Table 2 gives the results of spectral and morphological classifications.
As we discussed in section 3.1 we combine seven morphological types into three 
types, namely, galaxies with near or far companions (class 1 to 4),
interacting or merging galaxies (class 5,6) and isolated galaxies (class 0), and
assuming that the
isolate galaxies are at the last phase of merging, although there still exists the possibility that 
some of them are truly isolated galaxies and not at the end-state of merging.
These three new types construct a simple merger sequence. 
From table 2, we can see 
that the fraction of HII galaxies in VLIRGs decreases rapidly along the merger 
sequence, from 53\% (10/19) of class 1-4 to 39\% (16/41) of class 5-6 and
then to 23\% (3/13) of class 0.
 On the contrary, the fraction of AGN (LINER-like 
galaxies, Seyferts) increases dramatically. The corresponding 
fractions of LINER-like 
galaxies are 37\%, 37\% and 54\% and those of Seyfert 1s are 5\%, 7\% and 15\%,
respectively. Therefore, we can conclude that along the evolutionary
 sequence of 
interacting galaxy, the spectral properties of VLIRGs change
 from HII-like to AGN-like. 
In the last merger stage, they could evolve either to Seyfert 1's or LINERs. 

We have discussed in section 4.2 that quite a few VLIRGs are in loose or
compact groups 
of galaxies. For the ten confirmed VLIRGs in group of galaxies
in our sample, 8 are HII galaxies, one is a LINER-like galaxy and one is
an unclassified AGN. There are no 
Seyfert in these groups. 
The high proportion of HII galaxies indicates
 that VLIRGs in group of galaxies
are in the early evolution phase and the Seyferts or LINERs could also be formed
in the  more advanced stage of multiple-merger process.
Therefore, the existence of relatively large fraction 
$\ge$14\% (10/73) of VLIRGs in groups hints that merging is an 
important evolutionary process during galaxy formation.

\section{\bf Conclusion}

We have conducted a spectroscopic
survey for a complete sample of VLIRGs selected from the 2Jy redshift 
survey catalogue.
We have studied their optical properties and 
the spectral and morphological classifications
for these objects.
Important clues are found concerning the nature of VLIRGs and 
the possible evolution connection between starbursts and AGNs.



\begin{acknowledgements}
We would like to thank Dr. J.Y. Wei for his support for the project.
Many thanks are due 
to Profs. J.Y. Hu, J.H. Huang, H.J. Su and Dr. S. Mao for valuable
suggestions. We are grateful
to Prof. J.S. Chen and his group for their help in the data analysis.
Finally, we acknowledge Dr. M.A. Strauss for his valuable comments
on this paper. This work was partially supported by NSFC.

\end{acknowledgements}

\end{document}


\begin{center}
\small
\begin{tabular}{clcrccccccll}
\multicolumn{12}{c}{\large { \bf Table 1} {Reddenings, Observed and dereddened line ratios and spectral classifications}}\\
\multicolumn{12}{c}{}\\
\hline
\hline
  Name      &   &Galactic&$\rm H{\alpha}$ &    &[OIII]& [OI] & [NII]& [SII]  & C6563 &Spectral& Classification \\
	    &   &Extincion&----- &E(B-V)& -----& -----& -----& -----&-----&  Types  & in Veron\\
	    &   &E(B-V)&$\rm H{\beta}$ &       &$\rm H{\beta}$    &$\rm H{\alpha}$    &$\rm H{\alpha}$    &$\rm H{\alpha}$    &C4861&          & Catalog(1993)\\
     (1)    &   &  (2)  &  (3)  &  (4)  &  (5)  &  (6)  &  (7)  &  (8)  &  (9) &  (10) & (11) \\
\hline
  00189+3748&A  &  0.07 &  8.33 &  0.92 & -0.78:& -1.22 & -0.39 & -0.60 & 1.42 & LH          &\\
            &   &       &  3.10 &       & -0.83:& -1.17 & -0.39 & -0.63 &      &&\\
            &B  &  0.07 &  5.00 &  0.41 & -0.30 & -1.22;& -0.24 & -0.41 & 1.27 & LH          &\\
            &   &       &  3.10 &       & -0.32 & -1.19;& -0.25 & -0.43 &      &&\\
  00267+3016&A  &  0.04 & 14.29 &  1.48 &  0.20 & -1.05 & -0.21 & -0.39:& 1.00 & L           &\\
            &   &       &  3.10 &       &  0.12 & -0.97 & -0.21 & -0.43:&      &&\\
            &B  &  0.04 & 10.00 &  1.13 &  0.11 & -1.10 & -0.25 & -0.52:& 0.95 & LH          &\\
            &   &       &  3.10 &       &  0.06 & -1.03 & -0.26 & -0.56:&      &&\\
  00509+1225&   &  0.05 &  3.57 &  0.09 &  --   &  --   &  --   &  --   & 0.85 & S1          & S1\\
            &   &       &  3.10 &       &  --   &  --   &  --   &  --   &      &&\\
  01173+1405&   &  0.02 &  5.88 &  0.62 & -0.08 & -1.15 & -0.42 & -0.42 & 0.91:& LH          &\\
            &   &       &  3.10 &       & -0.12 & -1.12 & -0.42 & -0.44 &      &&\\
  01324+2138&   &  0.06 &  9.09 &  1.02 &  0.48 & -1.15 & -0.12 & -0.46:& 1.45:& LH          &\\
            &   &       &  3.10 &       &  0.43 & -1.10 & -0.12 & -0.49:&      &&\\
  01484+2220&   &  0.07 &  7.14 &  0.84 & -0.85:& -1.40 & -0.41:& -0.77 & 1.15 & H           &\\
            &   &       &  2.85 &       & -0.89:& -1.35 & -0.41:& -0.80 &      &&\\
  01572+0009&   &  0.00 &  2.86 &  0.00 &  0.05 & -2.00 & -0.85 & -1.52:& 0.80 & S1          & S1\\
            &   &       &  3.10 &       &  0.05 & -2.00 & -0.85 & -1.52:&      &&\\
  02071+3857&   &  0.04 & 33.33:&  2.42:&  0.12;& -1.70 & -0.35 & -0.62 & 1.19:& H           &\\
            &   &       &  2.85 &       &  0.00;& -1.57 & -0.36 & -0.69 &      &&\\
  02203+3158&A  &  0.06 & 14.29:&  1.46:&  0.46:& -1.22;& -0.14:& -0.51 & 1.12 & LH          & S?\\
            &   &       &  3.10 &       &  0.38:& -1.14;& -0.14:& -0.55 &      &&\\
            &B  &  0.06 & 10.00 &  1.19 &  --   & -2.00 & -0.41 & -0.62 & 0.94 & H           &\\
            &   &       &  2.85 &       &  --   & -1.93 & -0.41 & -0.66 &      &&\\
  02222+2159&   &  0.08 &  9.09;&  1.07;& -0.14;& -1.40 & -0.40 & -0.66:& 1.26:& H           &\\
            &   &       &  2.85 &       & -0.19;& -1.34 & -0.40 & -0.69:&      &&\\
  02248+2621&   &  0.06 &  5.88 &  0.66 & -0.39 & -1.70 & -0.17 & -0.68 & 0.74 & H           &\\
            &   &       &  2.85 &       & -0.42 & -1.66 & -0.18 & -0.70 &      &&\\
  02435+1253&   &  0.08 & 16.67;&  1.60;&  0.30;& -1.00:& -0.30:& -0.55:& 1.61:& LH          &\\
            &   &       &  3.10 &       &  0.22;& -0.91:& -0.31:& -0.61:&      &&\\
  02512+1446&A  &  0.07 &  7.14 &  0.85 & -0.24 & -1.52 & -0.35:& -0.55 & 0.98 & H           &\\
            &   &       &  2.85 &       & -0.29 & -1.48 & -0.35:& -0.58 &      &&\\
            &B  &  0.07 &  --   &  --   &  --   &  --   &  --   &  --   & 1.36 & O           &\\
            &   &       &  --   &       &  --   &  --   &  --   &  --   &      &&\\
  03117+4151&A  &  0.17 &  7.69 &  0.74 &  0.89 & -1.15 & -0.07 & -0.38 & 1.30 & S2          & S2\\
            &   &       &  3.10 &       &  0.85 & -1.11 & -0.07 & -0.40 &      &&\\
            &B  &  0.17 &  --   &  --   &  --   &  --   & -0.38 &  --   & 1.41 & H           &\\
            &   &       &  2.85 &       &  --   &  --   &  --   &  --   &      &&\\
            &C  &  0.17 &  --   &  --   &  --   &  --   &  --   &  --   & 1.27:& O           &\\
            &   &       &  --   &       &  --   &  --   &  --   &  --   &      &&\\
(05084+7936)&   &  0.09 & 10.00:&  1.16:& -0.22;& -1.30;& -0.39 & -0.62 & 1.06:& H           &\\
            &   &       &  2.85 &       & -0.28;& -1.24;& -0.39 & -0.66 &      &&\\
(05414+5840)&   &  0.29 &  --   &  --   &  --   &  --   &  0.26 & -0.21;& 1.53:& S?          &\\
            &   &       &  3.10 &       &  --   &  --   &  --   &  --   &      &&\\
(06538+4628)&   &  0.08 &  4.55 &  0.39 & -0.26 & -1.40 & -0.36 & -0.46 & 0.68:& H           &\\
            &   &       &  2.85 &       & -0.29 & -1.37 & -0.36 & -0.47 &      &&\\
  07063+2043&A  &  0.11 &  7.69 &  0.87 & -0.34 & -1.52 & -0.20 & -0.47 & 0.95:& H           &\\
            &   &       &  2.85 &       & -0.38 & -1.47 & -0.20 & -0.50 &      &&\\
(07062+2041)&B  &  0.11 &  4.35 &  0.30 & -0.36 & -1.40;& -0.40 & -0.51 & 0.82 & H           &\\
            &   &       &  2.85 &       & -0.38 & -1.38;& -0.40 & -0.52 &      &&\\
(07256+3355)&A  &  0.06 & 14.29 &  1.55 & -0.54:& -2.00;& -0.29 & -0.60 & 1.46:& H           &\\
            &   &       &  2.85 &       & -0.62:& -1.92;& -0.30 & -0.65 &      &&\\
            &B  &  0.06 &  3.57 &  0.17 &  --   & -1.52;& -0.39 & -0.44 & 0.79:& H           &\\
            &   &       &  2.85 &       &  --   & -1.51;& -0.39 & -0.45 &      &&\\
            &C  &  0.06 &  --   &  --   &  --   &  --   &  --   &  --   & 1.20 & O           &\\
            &   &       &  --   &       &  --   &  --   &  --   &  --   &      &&\\
\hline
\hline
\end{tabular}
\newpage
\begin{tabular}{clcrccccccll}
\multicolumn{12}{c}{\large { \bf Table 1} {--Continued}}\\
\multicolumn{12}{c}{}\\
\hline
\hline
\small
  Name      &   &Galactic&$\rm H{\alpha}$ &    &[OIII]& [OI] & [NII]& [SII]  & C6563 &Spectral& Classification \\
	    &   &Extincion&----- &E(B-V)& -----& -----& -----& -----&-----&  Types  & in Veron\\
	    &   &E(B-V)&$\rm H{\beta}$ &       &$\rm H{\beta}$    &$\rm H{\alpha}$    &$\rm H{\alpha}$    &$\rm H{\alpha}$    &C4861&          & Catalog(1993)\\
     (1)    &   &  (2)  &  (3)  &  (4)  &  (5)  &  (6)  &  (7)  &  (8)  &  (9) &  (10) & (11) \\
\hline
  08354+2555&   &  0.02 &  --   &  --   &  --   & -0.82;& -0.08 & -0.21:& 1.00 & S?    &\\
            &   &       &  3.10 &       &  --   &  --   &  --   &  --   &      &&\\
  08507+3520&A  &  0.01 &  --   &  --   &  --   &  --   & -0.02 & -0.14 & 1.07:& S?    &\\
            &   &       &  3.10 &       &  --   &  --   &  --   &  --   &      &&\\
            &B  &  0.01 &  --   &  --   &  --   & -1.05;& -0.21:& -0.30 & 1.05:& S?    &\\
            &   &       &  3.10 &       &  --   &  --   &  --   &  --   &      &&\\
            &C  &  0.01 &  --   &  --   &  --   & -1.05;& -0.17 & -0.15;& 0.88:& S?    &\\
            &   &       &  3.10 &       &  --   &  --   &  --   &  --   &      &&\\
(09047+1838)&   &  0.03 & 10.00 &  1.23 & -0.30:& -1.30 & -0.35 & -0.57 & 1.19 & H           &\\
            &   &       &  2.85 &       & -0.36:& -1.24 & -0.35 & -0.61 &      &&\\
  09126+4432&A  &  0.00 & 25.00;&  2.09;&  0.44;& -0.96 & -0.20 & -0.47:& 1.05 & LH          &\\
            &   &       &  3.10 &       &  0.34;& -0.85 & -0.21 & -0.53:&      &&\\
            &B  &  0.00 &  --   &  --   &  --   &  --   & -0.30 & -0.44;& 1.19:& H           &\\
            &   &       &  2.85 &       &  --   &  --   &  --   &  --   &      &&\\
  09168+3308&   &  0.00 &  4.00 &  0.34 & -0.44 & -1.30;& -0.36 & -0.54 & 0.86 & H           &\\
            &   &       &  2.85 &       & -0.46 & -1.28;& -0.36 & -0.55 &      &&\\
  09320+6134&   &  0.02 &  7.69:&  0.88:&  0.12:& -1.00;&  0.04 & -0.39 & 1.20 & L           & S\\
            &   &       &  3.10 &       &  0.07:& -0.95;&  0.03 & -0.41 &      &&\\
  09333+4841&A  &  0.01 &  9.09 &  1.15 & -0.04 & -1.30 & -0.24 & -0.48 & 0.92 & H           &\\
            &   &       &  2.85 &       & -0.10 & -1.24 & -0.24 & -0.52 &      &&\\
            &B  &  0.01 &  --   &  --   &  --   &  --   & -0.66 &  --   & 0.78:& H           &\\
            &   &       &  2.85 &       &  --   &  --   &  --   &  --   &      &&\\
(09517+6954)&   &  0.03 & 11.11 &  1.33 & -0.35 & -1.70 & -0.21 & -0.54 & 1.69 & H           &\\
            &   &       &  2.85 &       & -0.42 & -1.63 & -0.22 & -0.58 &      &&\\
  10203+5235&   &  0.00 &  5.88 &  0.64 &  0.17 & -1.10 & -0.21 & -0.40 & 0.99 & L           &\\
            &   &       &  3.10 &       &  0.14 & -1.06 & -0.21 & -0.42 &      &&\\
(10311+3507)&   &  0.00 &  5.26 &  0.61 & -0.68 & -1.30 & -0.40 & -0.57:& 0.90 & H           &\\
            &   &       &  2.85 &       & -0.71 & -1.27 & -0.40 & -0.59:&      &&\\
(10565+2448)&   &  0.00 &  8.33 &  0.99 & -0.04 & -1.05 & -0.19 & -0.52 & 0.80:& LH          &\\
            &   &       &  3.10 &       & -0.09 & -1.00 & -0.19 & -0.55 &      &&\\
  11231+1456&A  &  0.01 &  8.33 &  0.97 & -0.78:& -1.15 & -0.28 & -0.64 & 1.15:& LH          &\\
            &   &       &  3.10 &       & -0.83:& -1.10 & -0.28 & -0.67 &      &&\\
            &B  &  0.01 &  6.67 &  0.75 & -0.27:& -1.22 & -0.12 & -0.60 & 1.14:& LH          &\\
            &   &       &  3.10 &       & -0.31:& -1.18 & -0.12 & -0.63 &      &&\\
  11254+1126&   &  0.03 & 10.00 &  1.14 &  0.26 & -1.22 & -0.07 & -0.55 & 1.26 & LH          &\\
            &   &       &  3.10 &       &  0.20 & -1.16 & -0.07 & -0.59 &      &&\\
  11257+5850&A  &  0.00 &  2.94 &  0.00 & -0.05 & -1.52 & -0.47 & -0.60 & 0.69 & H           & H2\\
            &   &       &  2.85 &       & -0.05 & -1.52 & -0.47 & -0.60 &      &&\\
            &B  &  0.00 &  4.76 &  0.43 &  0.02 & -1.15 & -0.40 & -0.39 & 0.98 & LH          &\\
            &   &       &  3.10 &       &  0.00 & -1.13 & -0.40 & -0.40 &      &&\\
  11543+0124&   &  0.00 & 33.33;&  2.37;&  0.43;& -1.22 & -0.43 & -0.59:& 0.97:& LH          &\\
            &   &       &  3.10 &       &  0.31;& -1.10:& -0.44 & -0.66:&      &&\\
(12112+0305)&A  &  0.01 &  6.25:&  0.69:&  0.24:& -1.22;& -0.38 & -0.25:& 0.83:& LH          &\\
            &   &       &  3.10 &       &  0.21:& -1.19;& -0.38 & -0.27:&      &&\\
            &C  &  0.01 &  4.55:&  0.37:&  0.25:& -0.77;& -0.22 & -0.62:& 0.96:& LH          &\\
            &   &       &  3.10 &       &  0.23:& -0.75;& -0.22 & -0.63:&      &&\\
  12120+6838&A  &  0.01 &  7.69 &  0.90 & -0.03 & -1.15 & -0.24 & -0.32 & 1.13 & LH          &\\
            &   &       &  3.10 &       & -0.08 & -1.11 & -0.25 & -0.35 &      &&\\
            &C  &  0.01 &  6.25;&  0.69;& -0.20;& -1.22;& -0.38 & -0.28:& 0.89:& LH          &\\
            &   &       &  3.10 &       & -0.24;& -1.19;& -0.38 & -0.30:&      &&\\
            &D  &  0.01 &  4.55 &  0.46 & -0.09 & -1.40;& -0.48 & -0.44 & 0.68 & H           &\\
            &   &       &  2.85 &       & -0.11 & -1.37;& -0.48 & -0.46 &      &&\\
  12251+4026&   &  0.00 &  6.67 &  0.85 & -0.40 & -1.52 & -0.28 & -0.57 & 0.92 & H           &\\
            &   &       &  2.85 &       & -0.44 & -1.48 & -0.28 & -0.59 &      &&\\
  12265+0219&   &  0.00 &  1.43 &  0.00 & -0.70 &  --   & -1.40 &  --   & 0.50 & S1          & QSO\\
            &   &       &  3.10 &       & -0.70 &  --   & -1.40 &  --   &      &&\\
\hline
\hline
\end{tabular}
\newpage
\begin{tabular}{clcrccccccll}
\multicolumn{12}{c}{\large { \bf Table 1} {--Continued}}\\
\multicolumn{12}{c}{}\\
\hline
\hline
\small
  Name      &   &Galactic&$\rm H{\alpha}$ &    &[OIII]& [OI] & [NII]& [SII]  & C6563 &Spectral& Classification \\
	    &   &Extincion&----- &E(B-V)& -----& -----& -----& -----&-----&  Types  & in Veron\\
	    &   &E(B-V)&$\rm H{\beta}$ &       &$\rm H{\beta}$    &$\rm H{\alpha}$    &$\rm H{\alpha}$    &$\rm H{\alpha}$    &C4861&          & Catalog(1993)\\
     (1)    &   &  (2)  &  (3)  &  (4)  &  (5)  &  (6)  &  (7)  &  (8)  &  (9) &  (10) & (11) \\
\hline
  12323+1549&A  &  0.00 & 11.11;&  1.27;&  --   & -0.92;& -0.43 & -0.20 & 0.65:& LH          &\\
            &   &       &  3.10 &       &  --   & -0.86;& -0.44 & -0.24 &      &&\\
            &B  &  0.00 &  2.70 &  0.00 & -0.29 & -1.70;& -0.82:& -1.05 & 0.63:& S1          & S1\\
            &   &       &  3.10 &       & -0.29 & -1.70;& -0.82:& -1.05 &      &&\\
  12540+5708&   &  0.00 &  5.88 &  0.64 &  --   &  --   &  --   &  --   & 0.99 & S1          & S1\\
            &   &       &  3.10 &       &  --   &  --   &  --   &  --   &      &&\\
  12592+0436&   &  0.00 &  6.25;&  0.70;&  0.00;& -0.70;&  0.02 & -0.38:& 0.92 & L           &\\
            &   &       &  3.10 &       & -0.03;& -0.66;&  0.02 & -0.40:&      &&\\
  13136+6223&A  &  0.01 &  5.88 &  0.72 &  0.00 & -1.40 & -0.37 & -0.57 & 1.17 & H           &\\
            &   &       &  2.85 &       & -0.04 & -1.36 & -0.37 & -0.59 &      &&\\
            &B  &  0.01 &  9.09;&  1.07;&  0.30;& -1.22 & -0.24 & -0.28:& 1.39 & LH          &\\
            &   &       &  3.10 &       &  0.25;& -1.17 & -0.25 & -0.32:&      &&\\
  13183+3423&   &  0.00 & 33.33:&  2.37;&  0.48:& -0.92 & -0.28 & -0.47 & 1.04 & LH          &\\
            &   &       &  3.10 &       &  0.36:& -0.80 & -0.29 & -0.54 &      &&\\
  13299+1121&   &  0.00 &  5.26 &  0.61 & -0.16 & -1.52 & -0.38 & -0.55 & 0.98 & H           & H2\\
            &   &       &  2.85 &       & -0.19 & -1.49 & -0.38 & -0.57 &      &&\\
  13362+4831&A  &  0.00 &  4.35 &  0.34 &  0.18 & -0.92 & -0.19 & -0.32 & 1.05 & L           & S\\
            &   &       &  3.10 &       &  0.17 & -0.90 & -0.19 & -0.33 &      &&\\
            &B  &  0.00 &  5.88 &  0.64 &  0.77 & -1.40 & -0.25 & -0.39 & 1.01 & S2          & S2\\
            &   &       &  3.10 &       &  0.74 & -1.37 & -0.25 & -0.41 &      &&\\
  13373+0105&A  &  0.00 &  --   &  --   &  --   &  --   & -0.30 & -0.44:& 1.22 & LH          &\\
            &   &       &  3.10 &       &  --   &  --   &  --   &  --   &      &&\\
            &B  &  0.00 &  3.85 &  0.30 & -0.11 & -1.70;& -0.52 & -0.64:& 0.74 & H           &\\
            &   &       &  2.85 &       & -0.13 & -1.68;& -0.52 & -0.65:&      &&\\
  13428+5608&   &  0.00 & 11.11 &  1.28 &  0.90 & -0.96 & -0.06 & -0.28 & 1.30 & S2          & S2\\
            &   &       &  3.10 &       &  0.83 & -0.89 & -0.07 & -0.31 &      &&\\
  13458+1540&A  &  0.00 &  5.88 &  0.72 & -0.28 & -1.40;& -0.43 & -0.52 & 0.67 & H           &\\
            &   &       &  2.85 &       & -0.31 & -1.36;& -0.44 & -0.54 &      &&\\
            &B  &  0.00 &  4.17 &  0.38 & -0.20 & -1.22;& -0.35 & -0.54 & 0.79 & H           &\\
            &   &       &  2.85 &       & -0.22 & -1.20;& -0.35 & -0.55 &      &&\\
  13496+0221&A  &  0.01 & 12.50:&  1.47:& -0.30;&  --   & -0.21:& -0.66:& 1.01 & H           &\\
            &   &       &  2.85 &       & -0.37;&  --   & -0.22:& -0.70:&      &&\\
            &B  &  0.01 &  --   &  --   &  --   & -1.22;& -0.34 & -0.49:& 1.18:& H           &\\
            &   &       &  2.85 &       &  --   &  --   &  --   &  --   &      &&\\
            &C  &  0.01 &  --   &  --   &  --   &  --   & -0.59;&  --   & 0.88:& H           &\\
            &   &       &  2.85 &       &  --   &  --   &  --   &  --   &      &&\\
  13536+1836&   &  0.00 &  6.67 &  0.76 &  0.92 & -1.40 & -0.38:& -0.42 & 1.04 & S2          & S1\\
            &   &       &  3.10 &       &  0.88 & -1.36 & -0.38:& -0.44 &      &&\\
  14151+2705&A  &  0.00 & 25.00:&  2.09:&  0.30:& -1.70:& -0.14 & -0.46 & 0.98 & LH          & S2\\
            &   &       &  3.10 &       &  0.20:& -1.59:& -0.15 & -0.52 &      &&\\
            &B  &  0.00 &  --   &  --   &  --   &  --   & -0.35 &  --   & 1.20 & H           &\\
            &   &       &  2.85 &       &  --   &  --   &  --   &  --   &      &&\\
  14178+4927&   &  0.01 &  7.69 &  0.90 & -0.41;& -1.30;& -0.15 & -0.62 & 1.16 & LH          & H2\\
            &   &       &  3.10 &       & -0.46;& -1.25;& -0.15 & -0.65 &      &&\\
  14547+2448&A  &  0.03 &  8.33:&  0.96:&  0.28:& -1.00 & -0.10 & -0.26 & 1.55 & L           &\\
            &   &       &  3.10 &       &  0.23:& -0.95 & -0.10 & -0.29 &      &&\\
            &B  &  0.03 &  3.45 &  0.16 & -0.76;&  --   & -0.44:& -0.59;& 0.93:& H           &\\
            &   &       &  2.85 &       & -0.77;&  --   & -0.44:& -0.59;&      &&\\
  14568+4504&   &  0.01 &  5.26 &  0.52 &  0.18 & -1.30;& -0.17 & -0.52 & 1.05 & LH          &\\
            &   &       &  3.10 &       &  0.16 & -1.27;& -0.17 & -0.54 &      &&\\
  15107+0724&   &  0.01 &  --   &  --   &  --   &  --   & -0.29 & -0.42 & 1.04:& H           &\\
            &   &       &  2.85 &       &  --   &  --   &  --   &  --   &      &&\\
  15163+4255&   &  0.01 &  9.09 &  1.15 &  0.00 & -1.52 & -0.32 & -0.55 & 1.05 & H           &\\
            &   &       &  2.85 &       & -0.06 & -1.46 & -0.32 & -0.59 &      &&\\
\hline
\hline
\end{tabular}
\newpage
\begin{tabular}{clcrccccccll}
\multicolumn{12}{c}{\large { \bf Table 1} {--Continued}}\\
\multicolumn{12}{c}{}\\
\hline
\hline
\small
  Name      &   &Galactic&$\rm H{\alpha}$ &    &[OIII]& [OI] & [NII]& [SII]  & C6563 &Spectral& Classification \\
	    &   &Extincion&----- &E(B-V)& -----& -----& -----& -----&-----&  Types  & in Veron\\
	    &   &E(B-V)&$\rm H{\beta}$ &       &$\rm H{\beta}$    &$\rm H{\alpha}$    &$\rm H{\alpha}$    &$\rm H{\alpha}$    &C4861&          & Catalog(1993)\\
     (1)    &   &  (2)  &  (3)  &  (4)  &  (5)  &  (6)  &  (7)  &  (8)  &  (9) &  (10) & (11) \\
\hline
  15327+2340&   &  0.04 &  --   &  --   &  --   & -0.48;&  0.19 &  0.05:& 1.64:& S?    & S2\\
            &   &       &  3.10 &       &  --   &  --   &  --   &  --   &      &&\\
  15425+4114&A  &  0.01 &  5.56 &  0.65 &  0.09 & -1.30 & -0.44 & -0.49 & 0.72 & H           &\\
            &   &       &  2.85 &       &  0.05 & -1.27 & -0.45 & -0.51 &      &&\\
  15426+4116&B  &  0.01 &  2.94:&  0.00 &  --   & -0.59;& -0.22 & -0.64:& 0.94 & LH          &\\
            &   &       &  3.10 &       &  --   & -0.58;& -0.22 & -0.64:&      &&\\
  16104+5235&   &  0.00 &  5.56 &  0.67 & -0.35 & -2.00 & -0.35 & -0.70 & 0.87 & H           &\\
            &   &       &  2.85 &       & -0.38 & -1.97 & -0.35 & -0.72 &      &&\\
  16180+3753&   &  0.00 &  2.63 &  0.00 & -0.88;& -1.22;& -0.29 & -0.48:& 1.06 & H           &\\
            &   &       &  2.85 &       & -0.88;& -1.22;& -0.29 & -0.48:&      &&\\
  16284+0411&   &  0.06 & 10.00:&  1.10:& -0.52;& -1.10 & -0.28 & -0.66 & 1.42 & LH          &\\
            &   &       &  3.10 &       & -0.58;& -1.03 & -0.29 & -0.69 &      &&\\
  16504+0228&   &  0.07 & 16.67 &  1.61 &  0.30 & -0.46 &  0.06 &  0.03 & 2.54 & L           & S1\\
            &   &       &  3.10 &       &  0.22 & -0.37 &  0.05 & -0.02 &      &&\\
  16577+5900&A  &  0.00 &  2.56 &  0.00 &  --   & -0.89 & -0.29 & -0.36:& 1.29:& LH          &\\
            &   &       &  3.10 &       &  --   & -0.89 & -0.29 & -0.36:&      &&\\
            &B  &  0.00 &  9.09:&  1.07:&  0.00:& -1.15 & -0.28 & -0.41 & 1.24 & LH          &\\
            &   &       &  3.10 &       & -0.05:& -1.10 & -0.29 & -0.44 &      &&\\
  16589+0521&   &  0.07 & 16.67:&  1.70:& -0.30:& -1.52 & -0.42 & -0.52 & 2.59:& H           &\\
            &   &       &  2.85 &       & -0.39:& -1.43 & -0.43 & -0.57 &      &&\\
  17366+8646&   &  0.05 & 25.00:&  2.12:&  0.10:& -1.70;& -0.42 & -0.70 & 1.13 & H           &\\
            &   &       &  2.85 &       & -0.01:& -1.59;& -0.43 & -0.76 &      &&\\
  17392+3845&   &  0.03 &  7.14 &  0.89 & -0.45:& -1.70;& -0.31 & -0.55 & 1.35:& H           &\\
            &   &       &  2.85 &       & -0.49:& -1.65;& -0.31 & -0.58 &      &&\\
  17501+6825&A  &  0.05 &  5.88 &  0.68 & -0.23 & -1.40 & -0.33 & -0.60 & 1.09:& H           &\\
            &   &       &  2.85 &       & -0.27 & -1.36 & -0.33 & -0.62 &      &&\\
            &B  &  0.05 &  8.33 &  0.94 &  0.34 & -1.10 & -0.23 & -0.55 & 1.29:& LH          &\\
            &   &       &  3.10 &       &  0.29 & -1.04 & -0.23 & -0.58 &      &&\\
            &C  &  0.05 &  7.14;&  0.87;&  0.11;&  --   & -0.39 & -0.51:& 1.13:& H           &\\
            &   &       &  2.85 &       &  0.06;&  --   & -0.39 & -0.54:&      &&\\
            &D  &  0.05 &  5.00:&  0.43:& -0.52;& -1.05;& -0.26 & -0.33 & 1.09:& LH          &\\
            &   &       &  3.10 &       & -0.55;& -1.02;& -0.26 & -0.34 &      &&\\
  18525+5518&   &  0.07 &  9.09:&  1.09:&  0.00:& -1.30 & -0.31:& -0.55 & 1.22:& H           &\\
            &   &       &  2.85 &       & -0.06:& -1.24 & -0.31:& -0.59 &      &&\\
  18595+5048&   &  0.04 & 16.67;&  1.64;&  0.18;& -1.52;& -0.08 & -0.42 & 1.39 & LH          &\\
            &   &       &  3.10 &       &  0.09;& -1.44;& -0.09 & -0.47 &      &&\\
  19120+7320&A  &  0.09 & 11.11 &  1.18 &  0.65 & -1.52 & -0.24 & -0.57 & 1.37 & S2          &\\
            &   &       &  3.10 &       &  0.58 & -1.46 & -0.24 & -0.61 &      &&\\
            &B  &  0.09 &  9.09 &  1.07 & -1.04;& -1.70;& -0.35 & -0.66 & 1.62 & H           &\\
            &   &       &  2.85 &       & -1.10;& -1.64;& -0.35 & -0.69 &      &&\\
  20550+1656&   &  0.10 &  4.76 &  0.41 &  0.35 & -1.52 & -0.68 & -0.77 & 0.81 & H           &\\
            &   &       &  2.85 &       &  0.32 & -1.50 & -0.68 & -0.79 &      &&\\
  22388+3359&   &  0.09 &  --   &  --   &  --   & -1.30:& -0.16 & -0.47 & 1.20 & LH          &\\
            &   &       &  3.10 &       &  --   &  --   &  --   &  --   &      &&\\
  22501+2427&   &  0.05 &  4.35 &  0.37 & -0.41 & -1.52;& -0.24 & -0.92 & 0.61:& H           & H2\\
            &   &       &  2.85 &       & -0.43 & -1.50;& -0.25 & -0.93 &      &&\\
  23007+0836&A  &  0.03 &  3.33 &  0.04 & -0.15 & -2.00 & -0.64 & -0.96:& 1.00 & S1          & S1\\
            &   &       &  3.10 &       & -0.16 & -2.00 & -0.64 & -0.96:&      &&\\
            &B  &  0.03 &  --   &  --   &  --   & -0.96 & -0.30 & -0.36:& 0.98 & S?    &\\
            &   &       &  3.10 &       &  --   &  --   &  --   &  --   &      &&\\
  23024+1916&   &  0.03 &  9.09 &  1.04 &  0.04 & -1.30 & -0.21;& -0.42 & 1.61:& LH          &\\
            &   &       &  3.10 &       & -0.01 & -1.25 & -0.22;& -0.45 &      &&\\
  23135+2516&   &  0.07 &  9.09:&  1.01:&  0.48:& -1.15 &  0.00 & -0.48 & 1.23 & LH          &\\
            &   &       &  3.10 &       &  0.43:& -1.10 &  0.00 & -0.51 &      &&\\
\hline
\hline
\end{tabular}
\newpage
\begin{tabular}{clcrccccccll}
\multicolumn{12}{c}{\large { \bf Table 1} {--Continued}}\\
\multicolumn{12}{c}{}\\
\hline
\hline
\small
  Name      &   &Galactic&$\rm H{\alpha}$ &    &[OIII]& [OI] & [NII]& [SII]  & C6563 &Spectral& Classification \\
	    &   &Extincion&----- &E(B-V)& -----& -----& -----& -----&-----&  Types  & in Veron\\
	    &   &E(B-V)&$\rm H{\beta}$ &       &$\rm H{\beta}$    &$\rm H{\alpha}$    &$\rm H{\alpha}$    &$\rm H{\alpha}$    &C4861&          & Catalog(1993)\\
     (1)    &   &  (2)  &  (3)  &  (4)  &  (5)  &  (6)  &  (7)  &  (8)  &  (9) &  (10) & (11) \\
\hline
  23254+0830&A  &  0.03 & 33.33;&  2.43;&  --   & -1.40 & -0.31 & -0.36 & 1.13 & H           &\\
            &   &       &  2.85 &       &  --   & -1.27:& -0.32 & -0.43 &      &&\\
            &B  &  0.03 &  4.17 &  0.27 &  1.00 & -1.05 & -0.08 & -0.49 & 1.01 & S2          & S1\\
            &   &       &  3.10 &       &  0.99 & -1.03 & -0.08 & -0.50 &      &&\\
            &C  &  0.03 &  --   &  --   &  --   &  --   &  --   &  --   & 0.95 & O           &\\
            &   &       &  --   &       &  --   &  --   &  --   &  --   &      &&\\
            &D  &  0.03 &  --   &  --   &  --   &  --   &  --   &  --   & 1.00 & O           &\\
            &   &       &  --   &       &  --   &  --   &  --   &  --   &      &&\\
  23488+1949&A  &  0.03 & 10.00 &  1.22 & -0.52;& -1.70 & -0.42 & -0.72 & 1.08 & H           &\\
            &   &       &  2.85 &       & -0.58;& -1.63 & -0.43 & -0.76 &      &&\\
            &B  &  0.03 &  5.00 &  0.53 & -0.15 & -1.52 & -0.48:& -0.51:& 0.79 & H           &\\
            &   &       &  2.85 &       & -0.18 & -1.49 & -0.48:& -0.52:&      &&\\
            &C  &  0.03 &  7.14 &  0.80 & -0.85;& -1.15 & -0.40 & -0.66 & 0.88 & LH          &\\
            &   &       &  3.10 &       & -0.89;& -1.11 & -0.40 & -0.69 &      &&\\
  23488+2018&A  &  0.03 & 11.11 &  1.33 & -0.35:& -1.52 & -0.26 & -0.62 & 1.24 & H           &\\
            &   &       &  2.85 &       & -0.42:& -1.45 & -0.26 & -0.66 &      &&\\
            &B  &  0.03 &  3.57 &  0.20 &  0.23 & -1.22:& -0.89 & -0.57 & 0.83 & H           & \\
            &   &       &  2.85 &       &  0.22 & -1.21:& -0.89 & -0.57 &      &&\\
  23532+2513&A  &  0.04 &  --   &  --   &  --   &  --   & -0.34 &  --   & 1.29:& H           &\\
            &   &       &  2.85 &       &  --   &  --   &  --   &  --   &      &&\\
            &B  &  0.04 &  2.56:&  0.00:& -0.51 &  --   & -1.40 &  --   & 0.82:& S1          &\\
            &   &       &  3.10 &       & -0.51 &  --   & -1.40 &  --   &      &&\\
            &C  &  0.04 &  5.56 &  0.63 &  0.05 & -1.52 & -0.29 &  --   & 1.00:& H           &\\
            &   &       &  2.85 &       &  0.01 & -1.49 & -0.29 &  --   &      &&\\
  23594+3622&   &  0.09 & 33.33;&  2.28;&  1.00;& -0.74 &  0.30 & -0.14 & 1.15 & S2          &\\
            &   &       &  3.10 &       &  0.88;& -0.62:&  0.29 & -0.21 &      &&\\
\hline
\hline
\multicolumn{12}{l}{{\bf Notes}: }\\
\multicolumn{12}{l}{The sources in brackets are those not belong to the 73 complete samples. The uncertainty of value is about 15\%;}\\
\multicolumn{12}{l}{ The values with colon(:) and semicolon(;) have uncertainties 30\% and 50\% respectively.}\\
\multicolumn{12}{l}{Column ( 2) -- The Galactic extinction from machine-readable database (D. Burstein, 1993) }\\
\multicolumn{12}{l}{Column ( 3) -- Observed and theoretical $\rm H{\alpha}/H{\beta}$ values}\\
\multicolumn{12}{l}{Column ( 4) -- Color excesses determined from $\rm H{\alpha}/H{\beta}$ }\\
\multicolumn{12}{l}{Column ( 5) -- Logarithms of $\rm [OIII]\lambda5007/H{\beta}$  and dereddened ratio $\rm [OIII]\lambda5007/H{\beta}$  }\\
\multicolumn{12}{l}{Column ( 6) -- Logarithms of $\rm [OI]\lambda6300/H{\alpha}$  and dereddened ratio $\rm [OI]\lambda6300/H{\alpha}$  }\\
\multicolumn{12}{l}{Column ( 7) -- Logarithms of $\rm [NII]\lambda6584/H{\alpha}$  and dereddened ratio $\rm [NII]\lambda6584/H{\alpha}$}\\
\multicolumn{12}{l}{Column ( 8) -- Logarithms of $\rm [SII]\lambda6716+\lambda6731/H{\alpha}$  and dereddened ratio $\rm [SII]\lambda6716+\lambda6731/H{\alpha}$  }\\
\multicolumn{12}{l}{Column ( 9) -- Ratios of continuum fluxes at 6563$\rm \AA$ and 4861$\rm \AA$}\\
\multicolumn{12}{l}{Column (10) -- Spectral types determined from $\rm [OIII]/H{\beta}$, $\rm [OI]/H{\alpha}$, $\rm [NII]/H{\alpha}$  and $\rm [SII]/H{\alpha}$. {\it S1}: Seyfert 1; {\it S2}: Seyfert 2;} \\
\multicolumn{12}{l}{  {\it L}: LINER; {\it H}: HII galaxy; {\it S?}: Seyfert 2 or LINER; {\it LH}: mixture type with properties of both LINER and HII galaxy;} \\
\multicolumn{12}{l}{ {\it O}: other galaxy} \\
\multicolumn{12}{l}{Column (11) -- Spectral types from Verons' Catalog of quasars and active galactic nuclei(1993). {\it S1}: Seyfert 1; {\it S2}: Seyfert 2;}\\
\multicolumn{12}{l}{{\it H2}: HII galaxy; {\it S or S?}: Could be Seyfert or LINER.}\\
\end{tabular}
\end{center}